\documentclass{article}
\usepackage{amsmath}

\usepackage{arxiv}

\usepackage[utf8]{inputenc} 
\usepackage[T1]{fontenc}    
\usepackage{hyperref}       
\usepackage{url}            
\usepackage{booktabs}       
\usepackage{amsfonts}       
\usepackage{nicefrac}       
\usepackage{microtype}      
\usepackage{lipsum}		
\usepackage{graphicx}
\usepackage{subcaption}

\title{Generalized Finslerian Wormhole Models in $f(\mathcal{R},\mathcal{T})$ Gravity}

\author{Yashwanth B. R.$^1$, S. K. Narasimhamurthy$^{1*}$, and Z. Nekouee$^2$ \\
$^1$ Department of PG Studies and Research in Mathematics,
	Kuvempu University, Jnana Sahyadri, \\Shankaraghatta - 577 451, Shivamogga, Karnataka, India.\\
$^2$ School of Physics, Damghan University, Damghan, 3671641167, Iran.\\
	\texttt{Corresponding author$^*$: nmurthysk@gmail.com} \\
}

\hypersetup{pdfkeywords={Wormhole; Finsler space-time; $f(\mathcal{R}, \mathcal{T})$ gravity;  Energy conditions.}
}

\begin{document}
\maketitle

\begin{abstract}
This article explores wormhole solutions within the framework of Finsler geometry and the modified gravity theory. Modifications in gravitational theories, such as $f(\mathcal{R}, \mathcal{T})$ gravity, propose alternatives that potentially avoid the exotic requirements. We derive the field equations to examine the conditions for Finslerian wormhole existence and investigate geometrical and material characteristics of static wormholes using a polynomial shape function in Finslerian space-time. Furthermore, we address energy condition violations for different Finsler parameters graphically. We conclude that the proposed models, which assume a constant redshift function, satisfy the necessary geometric constraints and energy condition violations indicating the presence of exotic matter at the wormhole throat. The results are validated through analytical solutions and 3-D visualizations, contributing to the broader understanding of wormholes in Finsler-modified gravity contexts.
\end{abstract}


\section{Introduction}\label{sec.1}

A wormhole (WH) has a tube-like geometric structure with asymptotical flatness on either side. These are the hypothetical passages that connect the two distinct regions of space-time. For the first time in 1935, Einstein and Rosen provided mathematical evidence supporting the existence of these hypothetical passages, commonly known as Einstein - Rosen bridges \cite{Rosen}. The concept of WH  was first brought by Flamm \cite{Flamm}. Misner and Wheeler later made the first use of the phrase wormhole \cite{Wheel}. Although WH models in  General relativity (GR) call for the presence of exotic matter, now it is recognized that they can also occur in modified gravity theories with ordinary matter \cite{Bhawal, Maeda}. The radius of the throat of the WH can be thought of as either fixed in the case of static wormholes (SWHs) or variable in the case of non-static or cosmic WHs \cite{Rosen}.\\

It is well known that WHs offer a feasible technique for quick interstellar travel. They are the solution to Einstein's field equations and link two far-off cosmological locations. A traversable WH was first introduced by Morris and Thorne \cite{Thorne}. They examined spherically symmetric static objects using GR and demonstrated that they must violate energy conditions. Exotic matter in this energy-violating situation possesses physical characteristics that would violate laws of physics, which include particles with a negative mass. WHs are a tremendously intriguing issue in theoretical physics because of these exceptional characteristics \cite{Jusu, Rich, Hali}.\\

Modified gravitational theories have shown promise in reducing or eliminating the need for exotic matter. The emergence of dark universe theory has underscored the limitations of General Relativity (GR) and the necessity for modifications across vast distances and throughout the cosmos. This subject has spurred significant interest in modifying GR on cosmological scales, with researchers exploring various avenues, from dark energy models to higher-order curvature theories of gravity. $f(R)$ modified gravity, where $R$ represents the curvature scalar, has particularly drawn attention for its potential to address cosmic acceleration. Clifton et al. provided a comprehensive overview of advancements in modified gravity over recent decades \cite{BHATTI}.\\

T. Harko and colleagues presented an expansion of the $f(R)$ theories of gravity mentioned above by including the energy-momentum tensor trace $T$ as well as a general dependency on the Ricci scalar $R$ in the model's gravitational action, known as the $f(R, T)$ gravity \cite{Harko}. In fields including cosmology, thermodynamics, gravitational waves, and astrophysics, this alternative gravity hypothesis has been tested \cite{Momeni, Zubair}. Despite these attempts, WHs in $f(R, T)$ gravity theories still have a low information content. A specific instance of the static wormhole (SWH) geometry was explored, in which its redshift function is independent of both time and spatial coordinates \cite{Azizi, Waheed}. Since the dependency of $T$  in this modified $f(R, T)$ gravity appears from the inclusion of the quantum processes, it could be fascinating to examine more generic WH theories.\\

 Riemannian geometry forms the basis for describing space-time in Einstein's general relativity. However, given the challenges faced by general relativity, scientists have proposed modifications to the theory of gravity through advancements in differential geometry. Finsler geometry, as a generalization of Riemannian geometry, \cite{Bao2, Roxburgh} offers a framework where the line element depends on both space-time coordinates and tangent vectors. Recently, Finsler geometry has generated significant interest among physicists due to its potential to address several issues that Einstein's gravity cannot resolve. Nowadays, GR is used to explain a wide range of facts with remarkable precision. This theory faces few drawbacks in bulk and negligible scales \cite{Nekouee}. This Finsler framework holds GR without altering the dimensionality of space-time, making it better suitable for simultaneously explaining observers, gravity, and causal structures \cite{DESY}. Instead of changing the action of GR, Finslerian gravity theory is constructed by changing the geometric structure of the equations. The Finslerian space-time geometry is defined by a function on the tangent bundle rather than a base manifold. The investigation of extended dispersion relations within the Finsler space-time geometry yields results that are consistent with recent experimental observations \cite{Lorek}. Finsler-Lagrange-Hamilton geometry studies and Finsler cosmological models can be found in Ref. \cite{Vacaru, Vaca}. Also, Quantum gravity benefits greatly from Finslerian space-time geometry. Experimental findings and current conventional high-energy theories are compatible with Finsler-like gravity theories. Without relying on the dark matter hypothesis, Finsler geometry offers a superior tool to address the problems by the experimental findings of spiral galaxies, including their flat rotation curves \cite{Xli}.\\

In our investigation, we look at the wormhole solution through the perspectives of Finsler geometry \cite{Bao2}, offering an alternative context to general relativity. This study treats the four-velocity vector, a distinguishing feature of Riemannian geometry, as an independent variable. It is worth noting that in 1935 \cite{Cartan}, Cartan introduced self-consistent Finslerian models. Later, in 1950 \cite{Horvath}, Cartan d-connections were developed for Einstein-Finsler equations. This development prompted additional research into various Finsler geometry models applied to specific physics scenarios. While some studies used Finsler pseudo-Riemannian configurations, researchers had difficulty obtaining precise solutions in some cases.\\

In 2016, F. Rahaman et al. \cite{FRahman} presented WH models under the Finsler structure of space-time by considering the distinct options for shape function and energy density. This study was a follow-up to their last work \cite{Ray} that built a model for the compact stars based on Finsler geometry. Then, they developed their paper and investigated traversable wormholes by modifying some new solutions to wormholes in the framework of Finsler geometry supported by phantom energy \cite{FRahman1}. In 2022, H. M. Manjunatha studied WH models in $f(\mathcal{R})$ gravity with exponential shape function in the Finsler space-time geometry Ref. \cite{Manju} and discussed the WH solutions to the Einstein field equations in the view of anisotropic energy-momentum tensor by adopting Finslerian framework. Recently, M. Malligawad et al. \cite{Manju1} investigated the characteristics of WH models in the specific $f(\mathcal{R}, \mathcal{T})$ gravity. They analyzed WH solutions by adopting an exponential-type shape function. With continuous progress in the discovery of WH efforts, it is necessary to compile more predictions regarding their geometry of matter content. We aim to examine the wormhole model using the polynomial shape function in the perspective of the $f(\mathcal{R}, \mathcal{T})$ gravity theory in the Finslerian approach.\\

The paper is arranged as follows. In Section 2, we have discussed the Finslerian WH structure and its geometric formulations. In the Section 3, we derived  WH field equations in $f(\mathcal{R}, \mathcal{T})$ gravity. In Section 4, we constructed two distinct exact WH models from various hypotheses regarding their matter-geometry content, followed by discussions on the violation of energy conditions. Section 5 is devoted to a discussion and results. And finally, we conclude our article in Section 6.

\section{Finslerian Wormhole Structure}\label{sec.2}
One needs to instigate the metric to search the WH structure. Let us assume that the Finsler structure has the following shape \cite{Chang}.
\begin{equation}\label{Eq.1}
  \mathcal{F}^2 = \mathfrak{B}(r)y^ty^t - \mathcal{A}(r)y^ry^r - r^2 \bar{\mathcal{F}^2}(\theta,\phi,y^\theta,y^\phi).
\end{equation}

Here, $\mathfrak{B}(r) = e^{2a(r)},$ where $a(r)$ represents red shift function and $\mathcal{A}(r) = \left(1-\frac{\mathrm{b}(r)}{r}\right)^{-1}$, where $\mathrm{b}(r)$ is shape function of the WH, that must obey the following conditions.\\

\begin{itemize}
  \item [1.] The radial coordinate $r$ ranges from $r_{0}$ to $\infty$, with $r_{0}$ being the throat radius of the WH metric.
  \item [2.] At the throat, where $r= r_{0}$, the  shape function satisfies:

\begin{equation}\label{Eq.2}
  \mathrm{b}(r_{0})= r_{0},
\end{equation}
 For the region outside the throat i.e., $r> r_{0}$, the condition is:
\begin{equation}\label{Eq.3}
 1- \frac{\mathrm{b(r)}}{r}>0.
\end{equation}
  \item [3.] The  shape function must satisfy the flaring-out condition at the throat,
\begin{equation}\label{Eq.4}
 \mathrm{b}^{\prime}(r_{0})<1,
\end{equation}
where $^\prime$=$\frac{d}{dr}$.
  \item [4.] To ensure that the space-time geometry should be asymptotically flat it is required:
 \begin{equation}\label{Eq.5}
  \frac{\mathrm{b}(r)}{r}\rightarrow 0 \quad as\quad |r|\rightarrow \infty.
 \end{equation}
  \item [5.] $\mathfrak{{B}}(r)$ must be finite and non-vanishing at the throat $r_{0}$.\\
 With the reference \cite{Cataldo, aman}, we can consider $\mathfrak{{B}}(r) = k$, where $k$ is a constant, is used to achieve the anti-de Sitter and de Sitter asymptotic behaviors. Since a constant redshift function can be absorbed into the normalized time coordinate, we take $\mathfrak{B}(r)= 1$.\\

\end{itemize}
 In this study, $\bar{\mathcal{F}}$ is two-dimensional Finsler structure \cite{Wang}, and we take $\bar{\mathcal{F}^{2}}$ as:
\begin{equation}\label{Eq.6}
  \bar{\mathcal{F}}^2 = y^\theta y^\theta + \mathrm{f}(\theta, \phi)y^\phi y^\phi.
\end{equation}
So,
\begin{equation}\label{Eq.7}
  \bar{g}_{ij} = diag\left(1, \mathrm{f}(\theta, \phi)\right), \quad \quad
  \bar{g}^{ij} = diag\left(1, \frac{1}{\mathrm{f}(\theta, \phi)}\right),
\end{equation}
where $i,j = \theta, \phi.$\\
We can calculate the geodesic spray coefficients
\begin{equation}\label{Eq.8}
  G^\mu = \frac{1}{4}g^{\mu\nu}\left(\frac{\partial^2{\mathcal{F}}^2}{\partial{x^\lambda}\partial{y^\nu}}y^\lambda - \frac{\partial{\mathcal{F}}^2}{\partial{x^\nu}}\right),
\end{equation}
from $\bar{\mathcal{F}}^2$ as
\begin{eqnarray}
\bar{G^\theta} &=& - \frac{1}{4}\frac{\partial{\mathrm{f}}}{\partial{\theta}}y^\phi y^\phi,\\\label{Eq.9}
\bar{G}^\phi &=& \frac{1}{4\mathrm{f}}\left(2\frac{\partial{\mathrm{f}}}{\partial{\theta}}y^\phi y^\theta + \frac{\partial{\mathrm{f}}}{\partial{\phi}}y^\phi y^\phi\right),\label{Eq.10}
\end{eqnarray}
from these, we can obtain Ricci scalar $\overline{Ric}$:
\begin{equation}\label{Eq.11}
  \overline{Ric} = - \frac{1}{2\mathrm{f}}\frac{\partial^2{\mathrm{f}}}{\partial{\theta^2}} + \frac{1}{4{\mathrm{f}}^2}\left(\frac{\partial{\mathrm{f}}}{\partial{\theta}}\right)^2.
\end{equation}
Which might be a function of the $\theta$ constant.
For any constant $\eta$, we can have $\bar{{\mathcal{F}}^2}$ in three different cases:
\begin{eqnarray}\label{Eq.12}
 \bar{\mathcal{F}}^2 =\begin{cases}
  y^\theta y^\theta  + \mathcal{C}\sin^2(\sqrt{\eta}\theta)y^\phi y^\phi, & \eta > 0\\
   y^\theta y^\theta + \mathcal{C}\theta^2 (y^\phi)^2, & \eta = 0\\
   y^\theta y^\theta  + \mathcal{C}\sinh^2(\sqrt{-\eta}\theta)y^\phi y^\phi  & \eta < 0.
   \end{cases}
\end{eqnarray}

 We take $\mathcal{C} = 1$, which is trivial. Now the Eq. (\ref{Eq.1}) can be expressed as
\begin{eqnarray}\label{Eq.13}
  \mathcal{F}^2 &=& \mathfrak{B}(r)y^ty^t - \mathcal{A}(r)y^ry^r - r^2 y^\theta y^\theta - r^2 \sin^2\theta y^\phi y^\phi
         +r^2 \sin^2\theta y^\phi y^\phi - r^2\sin^2(\sqrt{\eta}\theta)y^\phi y^\phi.
\end{eqnarray}
Which is,
\begin{equation}\label{Eq.14}
  \mathcal{F}^2 = \alpha^2 + r^2\chi(\theta)y^\phi y^\phi,
\end{equation}
where $\alpha$ represents Riemannian metric and $\chi(\theta) = \sin^2\theta - \sin^2(\sqrt{\eta}\theta)$.
 Hence
 \begin{equation}\label{Eq.15}
   \mathcal{F} = \alpha\sqrt{1 + \frac{r^2 \chi(\theta)y^\phi y^\phi}{\alpha^2}}.
 \end{equation}
 We choose $b_\phi = r\sqrt{\chi(\theta)}$, we have
 \begin{equation}\label{Eq.16}
   \mathcal{F} = \alpha\phi(s),
 \end{equation}
where $\phi(s) = \sqrt{1 + s^2}$, $ s = \frac{b_\phi y^\phi}{\alpha} = \frac{\beta}{\alpha}$,  $b_\mu = (0, 0, 0, b_\phi)$,\quad $b_\phi y^\phi$ = $b_\mu y^\mu = \beta$\quad ($\beta$ is $1$-form).\\

Thus $\mathcal{F}$ is a  $(\alpha, \beta)$ type Finsler space.\\

Killing equation $\mathrm{K_V}(\mathcal{F}) = 0$ can be obtained in the Finsler space after the isometric transformations of the above Finsler structure Ref. \cite{Chang}:
\begin{equation}\label{Eq.17}
  \left(\phi(s) - s \frac{\partial \phi(s)}{\partial s}\right) \mathrm{K_V}(\alpha) + \frac{\partial \phi(s)}{\partial s} \mathrm{K_V}(\beta) = 0,
\end{equation}
\begin{eqnarray}\label{Eq.18}
 \mathrm{K_V}(\alpha) &=& \ \frac{1}{2\alpha}(\mathrm{V}_{\mu\mid\nu} + \mathrm{V}_{\nu\mid\mu})y^\mu y^\nu,\\
  \mathrm{K_V}(\beta) &=&  \left(\mathrm{V}^\mu \frac{\partial b_\nu}{\partial x^\mu} + b_\mu \frac{\partial \mathrm{V}^\mu}{\partial x^\nu}\right) y^\nu.
\end{eqnarray}
Here $"\mid "$ specifies the covariant derivative w.r.t. $\alpha$.\\
From the above, we have\\

$\mathrm{K_V}(\alpha) + s\mathrm{K_V}(\beta) = 0$ \quad or \quad $\alpha \mathrm{K_V}(\alpha) + \beta \mathrm{K_V}(\beta) = 0,$\\

which implies\\

\quad $\mathrm{K_V}(\alpha) = 0$ \quad and \quad $\mathrm{K_V}(\beta) = 0,$\\
                 {\centering or}\\
$\mathrm{V}_{\mu\mid\nu} + \mathrm{V}_{\nu\mid\mu} = 0$ \quad and \quad $\mathrm{V}^\mu \frac{\partial b_\nu}{\partial x^\mu} + b_\mu \frac{\partial \mathrm{V}^\mu}{\partial x^\nu} = 0$.\\

It is interesting to note that the first Killing equation is constrained by the second Killing equation. As a result, it is mainly responsible for shattering the Riemannian space's isometric symmetry.\\
The current Finsler metric space for $\bar{\mathcal{F}^2}$, considered as a quadric in $y^\theta$ and $y^\phi$, can be derived from the Riemannian manifold $(\mathcal{M}, g_{\mu\nu}(x))$ as follows:
$\mathcal{F}(x, y) = \sqrt{g_{\mu\nu}(x)y^\mu y^\nu}.$\\
It is important to note that this represents a semi-definite Finsler space. Consequently, the covariant derivative from the Riemannian space can be utilized.
We can then express the components of the Finsler metric (\ref{Eq.1}), where the metric
$\bar{g}_{ij} = diag(1, \sin^2\sqrt{\eta}\theta)$ \cite{Chowdhury}. Which is,
\begin{equation*}
g_{\mu\nu} = diag(\mathfrak{B}(r), -\mathcal{A}(r), -r^2, -r^2 \sin^2\sqrt{\eta}\theta).
\end{equation*}
Here, $\eta$ plays an important role in the outcoming field equations in the Finsler geometry and thereby influences the WH problem in our models.\\

 Eq. (\ref{Eq.1}) gives geodesic spray coefficients:\\
\begin{eqnarray}\label{Eq.19}
 G^t &=& 0,\\\nonumber
 G^r &=& \frac{r\mathrm{b}^\prime -\mathrm{b}}{4r(r-\mathrm{b})} y^r y^r - \frac{r-\mathrm{b}}{2}y^\theta y^\theta -\frac{r-\mathrm{b}}{2} \sin^2 (\sqrt{\eta}\theta) y^\phi y^\phi,\\ \nonumber
 G^\theta &=& \frac{1}{r} y^r y^\theta - \frac{\sqrt{\eta}}{2} \sin(\sqrt{\eta}\theta)\cos(\sqrt{\eta}\theta) y^\phi y^\phi,\\ \nonumber
G^\phi &=& \frac{1}{r}y^r y^\phi + \sqrt{\eta}\cot(\sqrt{\eta}\theta) y^\theta y^\phi. \nonumber
\end{eqnarray}
Now, we incorporate the geodesic spray coefficients from Eq. (\ref{Eq.19}) into Eq. (\ref{Eq.11}). Then we have
\begin{eqnarray}\label{Eq.20}
\mathcal{F}^2 Ric = \frac{r\mathrm{b}^\prime - \mathrm{b}}{r^2(r-\mathrm{b})} y^r y^r + \left(\eta - 1 + \frac{\mathrm{b}}{2r} + \frac{\mathrm{b}^\prime}{2}\right)y^\theta y^\theta + \left(\eta - 1 + \frac{\mathrm{b}}{2r} + \frac{\mathrm{b}^\prime}{2}\right)\sin^2 (\sqrt{\eta}\theta)y^\phi y^\phi.
\end{eqnarray}
Now we define scalar curvature $\mathcal{R}$ in  Finsler metric space as $ \mathcal{R} = g^{\mu\nu}Ric_{\mu\nu}$. Therefore, the modified Einstein tensors in the Finsler space-time can be derived as Ref. \cite{FRahman}
\begin{equation}\label{Eq.21}
  G_{\mu\nu} = Ric_{\mu\nu} - \frac{1}{2}\mathcal{R}g_{\mu\nu}.
\end{equation}
where $Ric_{\mu\nu}$ and $ G_{\mu\nu}$  are the Ricci tensor and Einstein tensors, respectively.
Chang and Li in Ref. \cite{Chang2} proved the covariant conservation of $G_{\mu\nu}$ in Finsler geometry
i.e., $G^{\mu}_{\nu|\mu}=0$.
The Bianchi identities in Finsler space coincide with those in Riemannian space, representing the covariant conservation of the Einstein tensor. When the current Finsler space reduces to the Riemannian space, the gravitational field equations can be obtained. According to Li et al. \cite{Chang1}, the gravitational field equations can also be discovered alternatively. They have demonstrated the covariance-preserving properties of the $G^{\mu}_{\nu}$ tensor w.r.t. the covariant derivative in Finsler space-time using the Chern-Rund connection.
It is important to note that the gravitational field equation in Finsler space is restricted to the base manifold of Finsler space \cite{Chang}, with the fiber coordinates $y^i$ set as the velocities of the cosmic components ( energy-momentum tensor velocities). Ref. \cite{Chang} shows that field equation can be obtained from the approximation work in \cite{Pfeifer}.
Pfeifer et al. studied the dynamics of gravitation in Finsler geometry using an action integral on the unit tangent bundle. The Ricci scalar depends solely on the Finsler metric structure $\mathcal{F}$ and is inconsiderate to the connection. Consequently, the gravitational field equation in Finsler space, being derived from the Ricci scalar, is inconsiderate to the connection. Thus, the Finsler gravitational field equation is given as follows (with $c=G=1$), \cite{DESY}:

\begin{equation}\label{Eq22.1}
G^{\mu}_{\nu}=8\pi_{\mathcal{F}}\mathcal{T}^{\mu}_{\nu},
\end{equation}
where $\mathcal{T}^{\mu}_{\nu}$ is the energy-momentum tensor and where $4\pi_\mathcal{F}$ represents the volume of the Finsler space-time structure $\bar{\mathcal{F}}$ .
The constituents of Finsler modified Einstein tensor is obtained from Eq. (\ref{Eq.21}) as follows Ref. \cite{FRahman1}:
\begin{eqnarray}\label{Eq.22}
 G^t_t &=& \frac{1}{r^2}(\mathrm{b}^\prime + \eta - 1),\\ \nonumber
 G^r_r &=& \frac{\mathrm{b}}{r^3} + \frac{1}{r^2}( \eta - 1),\\ \nonumber
 G^\theta_\theta &=& G^\phi_\phi = \frac{r\mathrm{b}^\prime - \mathrm{b}}{2r^3}. \nonumber
 \end{eqnarray}
  We consider the general anisotropic energy-momentum tensor Ref. \cite{aman} in the following form,
 \begin{equation}\label{Eq.23}
   \mathcal{T}^\mu_\nu = (\mathrm{\rho} + \mathrm{p}_t)u^\mu u_\nu + (\mathrm{p}_r - \mathrm{p}_t)\delta^\mu \delta_\nu - \mathrm{p}_tg^\mu_\nu,
 \end{equation}
In Finsler gravity, the total action explaining the gravitational interactions is structured similarly to Riemannian gravity but with notable distinctions owing to the unique features of Finsler geometry. In this context, the energy density  $\mathrm{\rho}= \mathrm{\rho}(r)$, radial pressure $\mathrm{p}_{r}= \mathrm{p}_{r}(r)$, and lateral pressure $\mathrm{p}_{t}= \mathrm{p}_{t}(r)$ (measured orthogonally to the radial direction) are the key components characterizing the gravitational system. Additionally, there is the four-velocity $u^\mu$ satisfying the condition $u^\mu u_\mu = 1$, and the space-like unit vector  $\delta^\mu$, with $\delta^\mu \delta_\mu = -1 $, considered in the radial direction.\\
In the formalism of Finsler gravity, the total action, as proposed by Harko \cite{Harko}, adopts a structure akin to that of Riemannian gravity. However, due to the distinct nature of Finsler geometry, there are significant differences in the formulation of the action, accounting for the specific geometric properties and gravitational interactions inherent in this framework. Stavrinos et al. \cite{PStavrinos} investigated modified gravity theories that modeled by the gravitational Lagrange density functionals $f(\mathcal{R}, \mathcal{T}, \mathcal{F})$ with generalized/ modified scalar curvature $\mathcal{R}$, trace of the matter field tensors $\mathcal{T}$ and the modified Finsler like generating function $\mathcal{F}$.
\begin{equation}\label{eq.86}
	\mathcal{S}=\frac{1}{16\pi}\int d^4x\sqrt{-g}f(\mathcal{R}, \mathcal{T})+\int d^4x \sqrt{-g}\mathcal{L}_m
\end{equation}
 where, $\mathcal{L}_m$ reprsents the matter Lagrangian density, $g$ is the determinant of the Finsler metric. Here, we concentrate on the functional form $f(\mathcal{R},\mathcal{T})= \mathcal{R}+2\lambda \mathcal{T}$, where $\mathcal{R}$ and $\mathcal{T}$ are scalar curvature and a function of the trace of the stress-energy tensor of matter, respectively.
 \begin{equation}\label{Eq86.1}
 \mathcal{T}_{\mu\nu}=\frac{-2}{\sqrt{-g}}\left[\frac{\partial(\sqrt{-g}\mathcal{L}_m)}{\partial g^{\mu\nu}}-\frac{\partial}{\partial x^\varsigma}\frac{\partial(\sqrt{-g}\mathcal{L}_m)}{\partial(\frac{\partial g^{\mu\nu}}{\partial x^\varsigma})}\right].
 \end{equation}
Here, we assume the units as $c=G=1$. By using Ref. \cite{Harko}, we suppose $\mathcal{L}_m$ depends only on the metric components and not on its derivatives, such that we get
\begin{equation}\label{Eq86.2}
\mathcal{T}_{\mu\nu}=g_{\mu\nu}\mathcal{L}_m-2\frac{\partial \mathcal{L}_m}{\partial g^{\mu\nu}}.
 \end{equation}
 By varying the action $\mathcal{S}$ of the gravitational field w.r.t. the metric tensor components of $g_{\mu\nu}$, we intend to derive the modified gravity field equation is close to the described method in Ref. \cite{saho}.
 \begin{eqnarray}\label{Eq86.3}
 \left(Ric_{\mu\nu}-\frac{1}{3}\mathcal{R}g_{\mu\nu}\right)&-&\frac{1}{6}f(\mathcal{R},\mathcal{T})g_{\mu\nu}=8\pi_{F}\left(\mathcal{T}_{\mu\nu}-\frac{1}{3}\mathcal{T}g_{\mu\nu}\right)
 -2\lambda\left(\mathcal{T}_{\mu\nu}-\frac{1}{3}\mathcal{T}g_{\mu\nu}\right)\nonumber\\&-&2\lambda\left(\Theta_{\mu\nu}-\frac{1}{3}\Theta g_{\mu\nu}\right),
 \end{eqnarray}
where $\Theta_{\mu\nu}=g^{\mu\nu}\frac{\partial \mathcal{T}_{\mu\nu}}{\partial g^{\mu\nu}}$ and $\Theta=\Theta^{\mu}_{\mu}$. In our current models, we let the matter Lagrangian $\mathcal{L}_m=\rho$. We discover the  modified gravitational field equation Finsler space-time as
\begin{equation}\label{Eq.24}
   G^\mu_\nu = (8\pi_\mathcal{F} + 2\lambda)\mathcal{T}^\mu_\nu + \lambda(2\rho + \mathcal{T})g^\mu_\nu,
\end{equation}
the Eq. (\ref{Eq.24}) reduce to Eq. (\ref{Eq22.1}) when $\lambda=0$.
To obtain the gravitation field equations, we assume an anisotropic fluid obeying the matter content of the following form,
  \begin{equation}\label{Eq.25}
    \mathcal{T}^{\mu}_{\nu}= diag(\mathrm{\rho}, -\mathrm{p}_{r}, -\mathrm{p}_{t}, -\mathrm{p}_{t}).
  \end{equation}
  The trace $\mathcal{T}$ appears to be $\mathcal{T}= \mathrm{\rho}- \mathrm{p}_{r}- 2\mathrm{p}_{t}$.

\section{Wormhole Field Equations in $f(\mathcal{R}, \mathcal{T})$ Gravity}\label{sec.3}
The constituents of the gravitational field  equations (\ref{Eq.24}) for the considered metric (\ref{Eq.13}) with Eq. (\ref{Eq.25}) are
\begin{eqnarray}\label{65}
G^t_t &=& (8\pi_\mathcal{F} + \lambda)\mathrm{\rho} - \lambda(\mathrm{p}_r + 2\mathrm{p}_t): \hspace{1.5cm} \frac{1}{r^2}(\mathrm{b}^\prime + \eta - 1) = (8\pi_\mathcal{F} + \lambda)\mathrm{\rho} - \lambda(\mathrm{p}_r + 2\mathrm{p}_t), \nonumber\\
G^r_r &=& \lambda\mathrm{\rho} + (8\pi_\mathcal{F} + 3\pi)\mathrm{p}_r + 2\lambda \mathrm{p}_t:  \hspace{1.4cm} -\frac{\mathrm{b}}{r^3} - \frac{1}{r^2}(\eta - 1) = \lambda\mathrm{\rho} + (8\pi_\mathcal{F} + 3\pi)\mathrm{p}_r + 2\lambda \mathrm{p}_t, \nonumber\\
G^\theta_\theta &=& G^\phi_\phi = \lambda\mathrm{\rho} + \lambda \mathrm{p}_r + (8\pi_\mathcal{F} + \lambda)\mathrm{p}_t: \hspace{0.7cm} \frac{\mathrm{b} - r\mathrm{b}^\prime }{2r^3} = \lambda\mathrm{\rho} + \lambda \mathrm{p}_r + (8\pi_\mathcal{F} + \lambda)\mathrm{p}_t. \nonumber
\end{eqnarray}
The above set of field equations concede the solutions
 \begin{eqnarray}
 \mathrm{\rho} &=& \frac{\mathrm{b}^{\prime}+(\eta - 1)}{r^{2}(8\pi_\mathcal{F} + 2\lambda)},\\\label{Eq.27}
  \mathrm{p}_{r} &=& - \frac{\mathrm{b} r+ r(\eta - 1)}{r^{3}(8\pi_\mathcal{F} + 2\lambda )},\\\label{Eq.28}
  \mathrm{p}_{t} &=& \frac{\mathrm{b}- \mathrm{b}^{\prime}r}{2r^{3}(8\pi_\mathcal{F} + 2\lambda )}.\label{Eq.29}
  \end{eqnarray}
 Field equations derived from the Ricci scalar remain independent of the connections and are thus insensitive to their variations. Additionally, these field equations could be deduced through a Lagrangian perspective. Notably, the contribution of $\eta$, representing the beta component of the fundamental function in Finsler space, is evident in these field equations, providing the distinctive Finslerian influence. In studying gravitational field equations within general relativity and gravitation, it's essential to consider the approach of the Cartesian connection. This method is more conventional and crucial as it maintains an angle between $2$ vectors passing along the geodesics and preserves their norm. This aspect is fundamental in deriving Einstein's gravitational equations. While we have chosen to avoid this approach in our study, it remains a feasible option for exploration.
\section{Wormhole Models}\label{sec.4}
In this section, we'll develop WH models based on the different hypotheses regarding their matter content.
\subsection{Model 1}\label{sec.4.1}
Here, we suppose the pressure $\mathrm{p}_t$ and $\mathrm{p}_r$ can be related as
 \begin{equation}\label{Eq.30}
  \mathrm{p}_{t}= \mathrm{n}\mathrm{p}_{r},
 \end{equation}
with $\mathrm{n}$ as an arbitrary constant. Such relations are taken in \cite{Lobo}, for instance.\\
Using Eq. (\ref{Eq.30}) in Eqs. (\ref{Eq.28}) and (\ref{Eq.29}), we get
\begin{equation}\label{Eq.31}
  \mathrm{b(r)}= (1- \eta)r + \mathrm{A}r^{1+2\mathrm{n}},
\end{equation}
where $\mathrm{A} =$ integral constant. The above Eq. (\ref{Eq.31})  is stable when $\mathrm{n}$ is negative and $\mathrm{A}$ is positive because of the asymptotic flatness of the metric and also flaring-out condition is satisfied.\\

\begin{figure}[hpbt]
\begin{center}
\includegraphics[scale=0.42]{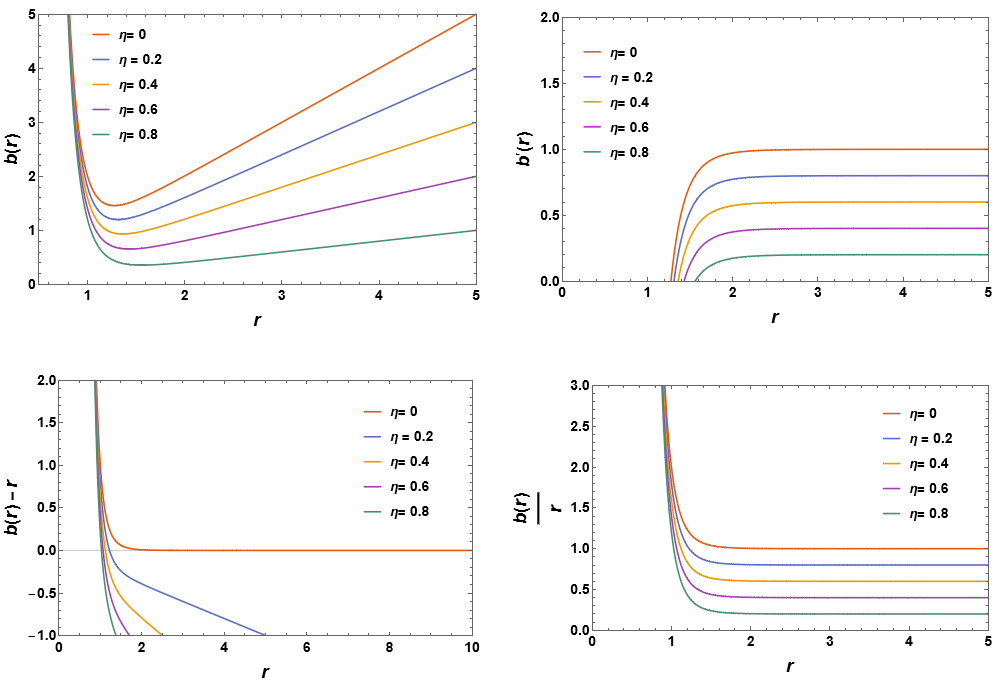}
 \caption{\label{fig8} Characteristic graphs of the shape function ($\mathrm{b(r)}$) described by Model 1, for $\mathrm{A}=1, \mathrm{n} = -4.$}
\end{center}
\end{figure}

The shape function is plotted w.r.t. $r$ in Fig. \ref{fig8} with $\mathrm{n}= -4$ and $\mathrm{A} =1$. From such a figure, all the basic WH conditions are satisfied.\\

\begin{figure}[hpbt]
\begin{center}
\includegraphics[scale=0.4]{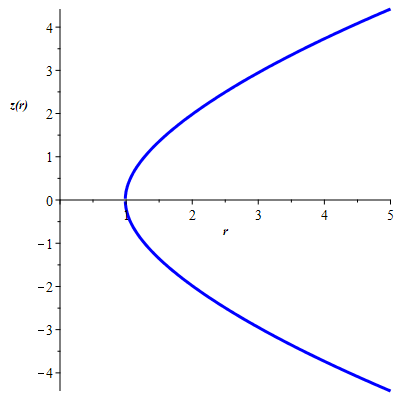}
 \caption{\label{fig22}Nature of the shape function Eq. (\ref{Eq.31}), at $r_0 = 1.$}
\end{center}
\end{figure}


At the throat of the wormhole i.e., $r= r_{0}$, then the Eq. (\ref{Eq.2}) gives
\begin{equation}\label{Eq.32}
  \mathrm{A} = \frac{\eta}{r^{2 \mathrm{n}}_{0}}.
\end{equation}
And we have energy density,
\begin{equation}\label{Eq.33}
  \rho= \frac{(1+2\mathrm{n})\mathrm{A}}{8\pi_\mathcal{F}+ 2\lambda} r^{2(\mathrm{n}-1)},
\end{equation}
radial pressure and lateral pressure
\begin{equation}\label{Eq.34}
  \mathrm{p}_r= \frac{-\mathrm{A}}{8\pi_\mathcal{F} + 2\lambda} r^{2(\mathrm{n}-1)},
\end{equation}
and
\begin{equation}\label{Eq.35}
  \mathrm{p}_t= \frac{-\mathrm{nA}}{8\pi_\mathcal{F} + 2\lambda} r^{2(\mathrm{n}-1)}.
\end{equation}
From Eqs. (\ref{Eq.33})-(\ref{Eq.35}) we have
\begin{equation}\label{Eq.36}
  \mathrm{\rho}+ \mathrm{p}_r = \frac{\mathrm{nA}}{4\pi_\mathcal{F} + \lambda}r^{2(\mathrm{n}-1)},
\end{equation}

\begin{equation}\label{Eq.37}
  \mathrm{\rho}+\mathrm{p}_t = \frac{\mathrm{A}(1+\mathrm{n})}{8\pi_\mathcal{F} + 2\lambda}r^{2(\mathrm{n}-1)}.
\end{equation}
The presence of the exotic matter inside the WH violates energy conditions. Specifically, the null energy condition (NEC) is violated by the energy-momentum tensor at the throat of the WH \cite{Hochberg}. We can note from the Eq. (\ref{Eq.36}) violation of the NEC, i.e., $\rho+ \mathrm{p}_r \leq 0$, that intimates $\lambda > -4\pi$.

The dominant energy condition (DEC) is given by,
\begin{equation}\label{Eq.38}
   \mathrm{\rho}- \mathrm{p}_r = \frac{\mathrm{A}(1+\mathrm{n})}{4\pi_\mathcal{F} + \lambda}r^{2(\mathrm{n}-1)},
\end{equation}
\begin{equation}\label{Eq.39}
   \mathrm{\rho}- \mathrm{p}_t = \frac{\mathrm{A}(1+3\mathrm{n})}{8\pi_\mathcal{F} + 2\lambda}r^{2(\mathrm{n}-1)}.
\end{equation}

\subsection{Model 2}\label{sec.4.2}
In this model, we consider the matter along with the equation of state (EoS)
\begin{equation}\label{Eq.40}
  \mathrm{p}_r + \omega(r)\mathrm{\rho} = 0,
\end{equation}
is  stuffing WH, where $\omega(r)> 0$, for the radial coordinate. A similar EoS with the varying parameter $\omega (r)$ is considered in the Ref. \cite{Raha}, for instance.
Taking Eq. (\ref{Eq.40}) into account, from  Eqs. (\ref{Eq.27}) and (\ref{Eq.28}) we can get
\begin{equation}\label{Eq.41}
  \omega(r)= \frac{\mathrm{b}+r(\eta-1)}{r(\mathrm{b}^\prime + (\eta-1))}.
\end{equation}

Further, we shall enquire about two cases for Eq. (\ref{Eq.41}), as follows in \cite{Raha}.\\
\subsubsection{Case  I:}\label{sec.4.21}
\begin{figure}[hpbt]
\begin{center}
\includegraphics[scale=0.37]{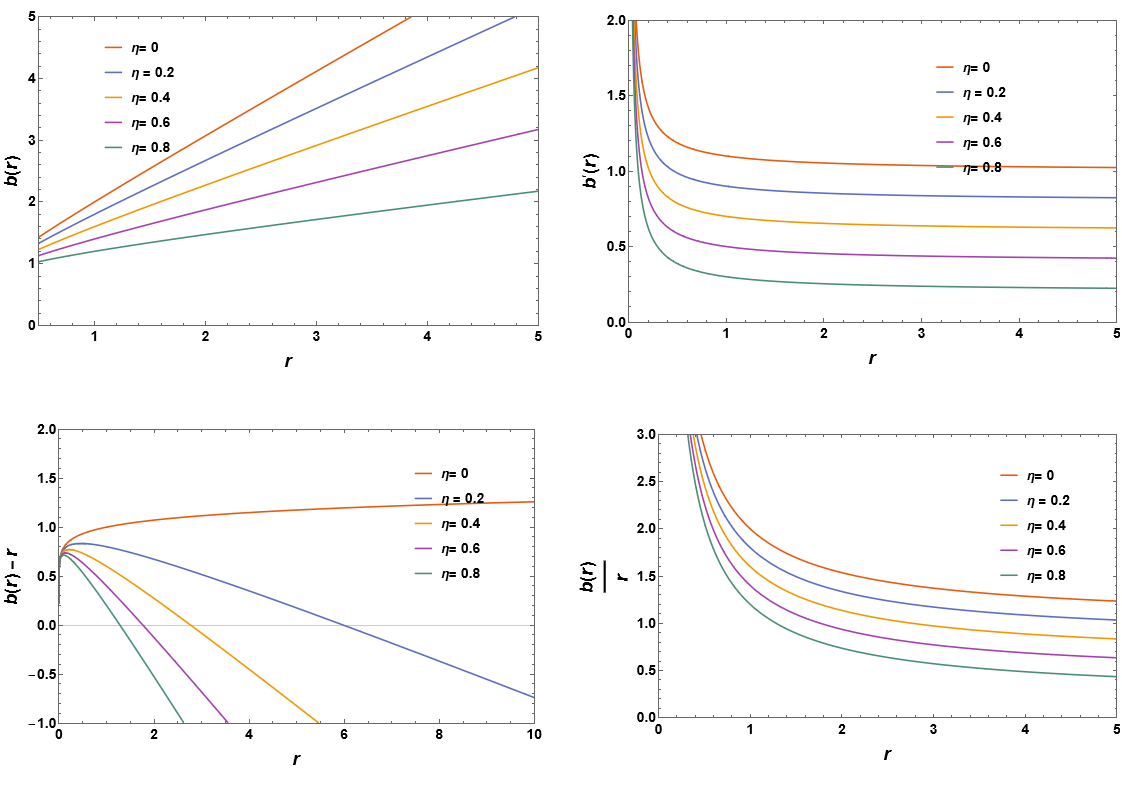}
 \caption{\label{fig9} Characteristic graphs of the shape function ($\mathrm{b(r)}$) described by Model 2 (case I), for the values  $\mathrm{b}_0 = 1$ and $\omega = 10$.}
\end{center}
\end{figure}

$ \omega(r)= \omega \quad(\textrm{a constant})$.\\
Now the Eq. (\ref{Eq.41}) becomes,
\begin{equation}\label{Eq.42}
  \mathrm{b(r)}= (1-\eta)r + \mathrm{b}_0r^{1/\omega},
\end{equation}
Given that, $\mathrm{b}_0$ = integral constant and considering the asymptotical flatness of the metric, Eq. (\ref{Eq.42}) remains stable when $\omega>1$. By selecting specific parameter values, we plot $\mathrm{b(r)}$  in Fig. \ref{fig9}. It can be observed that $r>r_0$ and $\mathrm{b(r)}-r<0$ is a crucial condition to fulfill for shape function. Additionally, $\mathrm{b(r)}-r$ is a decreasing function for $r>r_0$, which satisfies the flaring-out condition.

 At the throat of the WH i.e., $r=r_0$, that implies $\mathrm{b(r)}= r_0 $, which intimates
 \begin{equation}\label{Eq.43}
    r_0 = \left(\frac{\mathrm{b}_0}{\eta}\right)^{\frac{\omega}{\omega-1}},
 \end{equation}
utilizing the Eq. (\ref{Eq.42}) in Eqs. (\ref{Eq.27})-(\ref{Eq.29}), we have
\begin{eqnarray}
\mathrm{\rho} &=& \frac{-(\eta-1)r^{-2}\omega + \mathrm{b}_0 r^{\frac{1-3\omega}{\omega}}}{\omega(8\pi_\mathcal{F} + 2\lambda)},\\\label{Eq.44}
\mathrm{p}_r &=& \frac{(\eta-1)r^{-2} - \mathrm{b}_0r^{\frac{1-3\omega}{\omega}}}{8\pi_\mathcal{F} + 2\lambda},\\\label{Eq.45}
\mathrm{p}_t &=& \frac{\mathrm{b}_0(\omega-1)r^{\frac{1-3\omega}{\omega}}}{2\omega(8\pi_\mathcal{F} + 2\lambda)}.\label{Eq.46}
  \end{eqnarray}

\begin{figure}[hpbt]
\begin{center}
\mbox{{\includegraphics[scale=0.52]{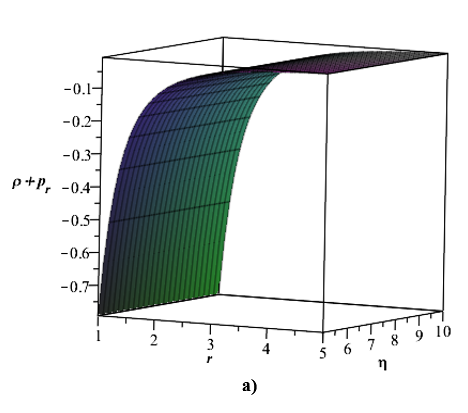}}
{\includegraphics[scale=0.5]{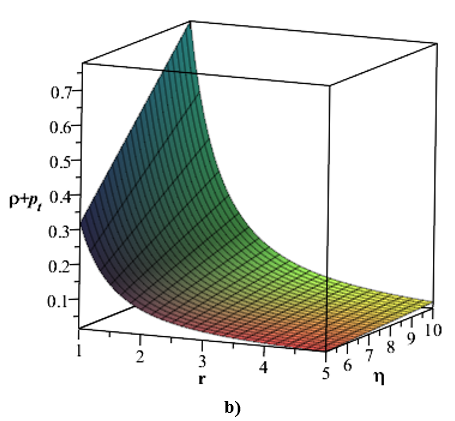}}}
 \caption{\label{fig15} a) Violation  of the NEC, $\mathrm{\rho}+ \mathrm{p}_r \leq 0$, for $\mathrm{b}_0 = 1$, and $\lambda = -12$. b) \label{fig1} Validity region of the NEC, $\mathrm{\rho}+ \mathrm{p}_t \geq 0$, for $\mathrm{b}_0 = 1$, and $\lambda = -12$.}
\end{center}
\end{figure}

NEC is given by,

\begin{eqnarray}
\mathrm{\rho} + \mathrm{p}_r &=& \frac{ \mathrm{b}_0(1- \omega)r^{\frac{1-3\omega}{\omega}}}{\omega(8\pi_\mathcal{F} + 2\lambda)},\\\label{Eq.47}
\mathrm{\rho} + \mathrm{p}_t &=& \frac{ -2(\eta-1)r^{-2}\omega + \mathrm{b}_0(1- \omega)r^{\frac{1-3\omega}{\omega}}}{2\omega(8\pi_\mathcal{F} + 2\lambda)},\label{Eq.48}
\end{eqnarray}

from the Eq. (\ref{Eq.47}) NEC is the violated for the values $\mathrm{b}_0>0$ and $\lambda> -4\pi$.
NEC for the present case can be seen in Fig. \ref{fig1}.

DEC is given by,

\begin{eqnarray}
\mathrm{\rho} - \mathrm{p}_r &=& \frac{ -2(\eta-1)r^{-2}\omega + \mathrm{b}_0(1+ \omega)r^{\frac{1-3\omega}{\omega}}}{\omega(8\pi_\mathcal{F} + 2\lambda)},\\\label{Eq.49}
\mathrm{\rho} - \mathrm{p}_t &=& \frac{ -2(\eta-1)r^{-2}\omega + \mathrm{ b}_0(3- \omega)r^{\frac{1-3\omega}{\omega}}}{2\omega(8\pi_\mathcal{F} + 2\lambda)}.\label{Eq.50}
\end{eqnarray}

\begin{figure}[hpbt]
\begin{center}
\mbox{{\includegraphics[scale=0.52]{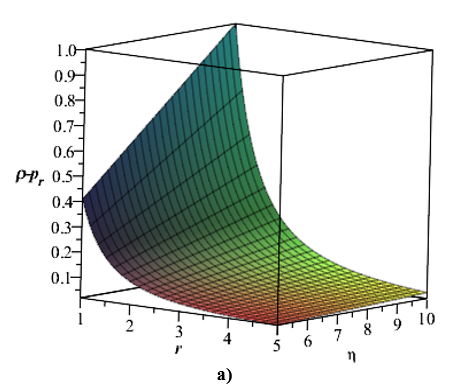}}
{\includegraphics[scale=0.5]{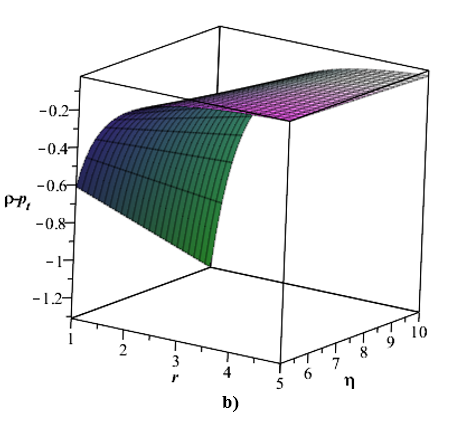}}}
 \caption{
\label{fig2}a) Validity region of the DEC, $\mathrm{\rho} \geq |\mathrm{p}_r|$, for $\mathrm{b}_0 = 1$, and $\lambda = -12$. \label{fig3}b) Validity region of the DEC, $\mathrm{\rho} \geq |\mathrm{p}_t|$, for $\mathrm{b}_0 = 1$, and $\lambda = -12$.
}

\end{center}
\end{figure}

DEC for this case can be seen in Fig. \ref{fig2}.

\
\
\\
\subsubsection{Case II:}\label{sec.4.22}
 $\omega(r) = \mathcal{B}r^\mathrm{m}$

In this case, we assume $\omega(r) = \mathcal{B}r^\mathrm{m}$, where, $\mathcal{B}$ and $\mathrm{m}$ are positive constants. From the Eq. (\ref{Eq.41}), $\mathrm{b}(r)$ is obtained as
\begin{equation}\label{Eq.51}
  \mathrm{b(r)} = -r(\eta-1)+ \exp{\left(c- \frac{1}{\mathrm{m}\mathcal{B}r^\mathrm{m}}\right)},
\end{equation}
where $c$ is the integration constant.

At the WH throat, we can have
\begin{equation}\label{Eq.52}
  c = \ln(\eta r_0) + \frac{1}{\mathrm{m}\mathcal{B}r^\mathrm{m}_0},
\end{equation}

so Eq. (\ref{Eq.51}) implies
\begin{equation}\label{Eq.53}
  \mathrm{b(r)}= -r(\eta-1)+ \exp{\left(\ln(\eta r_0) + \frac{1}{\mathrm{m}\mathcal{B}}\left(\frac{1}{r^\mathrm{m}_0}- \frac{1}{r^\mathrm{m}}\right)\right)},
\end{equation}

\begin{figure}[hpbt]
\begin{center}
\includegraphics[scale=0.45]{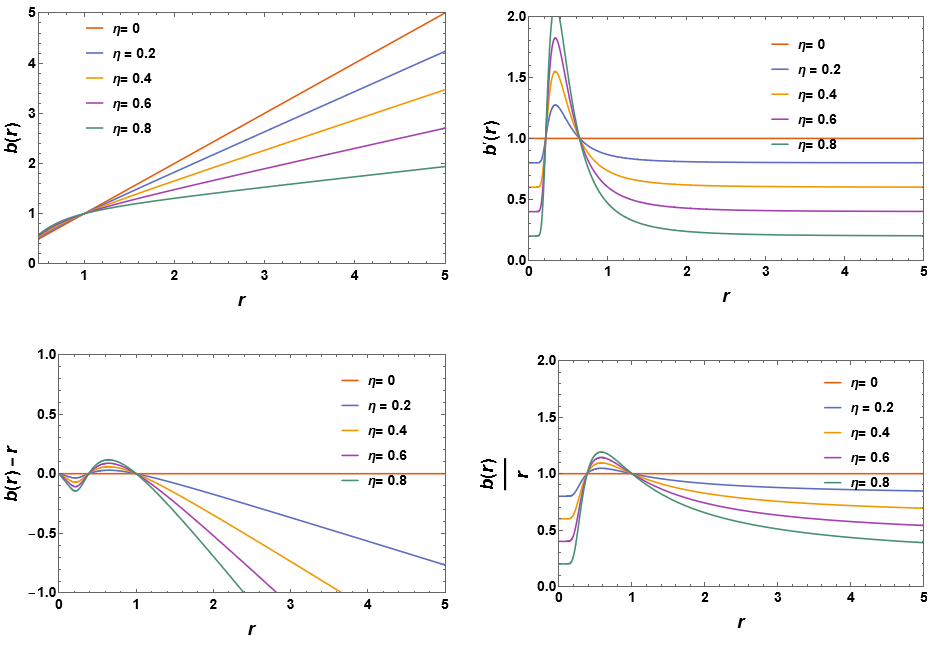}
 \caption{\label{fig10} Characteristic graphs of the shape function ($\mathrm{b(r)}$) described by Model 2 (case II), for the values $  r_0 = 1, \mathrm{m} = 2$, and $\mathcal{B} = 3$.}
\end{center}
\end{figure}

Since $\mathrm{m}>0$ and $\omega>1$, we have $r>r_0>(\frac{1}{\mathcal{B}})^{(\frac{1}{\mathrm{m}})}$, which satisfies the asymptotic flatness. Therefore, our assumption of EoS is true. From the Fig. \ref{fig10}, we observe, when $r>r_0$, $\mathrm{b}(r) - r$ is a decreasing function of $r$ for \(r \geq r_0\) and $\mathrm{b}^\prime(r_0)<1$, that satisfies the metric flaring - out condition. Thus, the plotted shape function $\mathrm{b}(r)$ indeed represents a structure of WH.

Using Eq. (\ref{Eq.51}) into Eqs. (\ref{Eq.27})-(\ref{Eq.29}), we get
\begin{eqnarray}
  \mathrm{\rho}&=& \frac{-(\eta-1) \mathcal{B}r^{(\mathrm{m}+1)} + \exp(c- \frac{1}{\mathrm{m}\mathcal{B}r^\mathrm{m}})}{\mathcal{B}r^{(\mathrm{m}+3)}(8\pi_\mathcal{F} +2\lambda)},\\\label{Eq.54}
\mathrm{p}_r &=& \frac{r (\eta-1)- \exp(c- \frac{1}{\mathrm{m}\mathcal{B}r^\mathrm{m}})}{r^{3}(8\pi_\mathcal{F} +2\lambda)},\\\label{Eq.55}
\mathrm{p}_t &=& \frac{(\mathcal{B}r^\mathrm{m} -1) \exp(c- \frac{1}{\mathrm{m}\mathcal{B}r^\mathrm{m}})}{2\mathcal{B}r^{(\mathrm{m}+3)}(8\pi_\mathcal{F} +2\lambda)}.\label{Eq.56}
\end{eqnarray}

NEC for this model is given by

\begin{eqnarray}
\mathrm{\rho}+ \mathrm{p}_r &=& \frac{\exp(c- \frac{1}{\mathrm{m}\mathcal{B}r^\mathrm{m}})(1- \mathrm{B}r^\mathrm{m})}{\mathrm{B}r^{(\mathrm{m}+3)}(8\pi_\mathcal{F} +2\lambda)},\\\label{Eq.57}
\mathrm{\rho}+ \mathrm{p}_t &=& \frac{-2(\eta-1)\mathcal{B}r^{(\mathrm{m}+1)} + \exp(c- \frac{1}{\mathrm{m}\mathcal{B}r^\mathrm{m}})(\mathcal{B}r^\mathrm{m}+1)}{2(\mathcal{B}r^{(\mathrm{m}+3)})(8\pi_\mathcal{F} +2\lambda)}.\label{Eq.58}
\end{eqnarray}

\begin{figure}[hpbt]
\begin{center}
\mbox{{\includegraphics[scale=0.52]{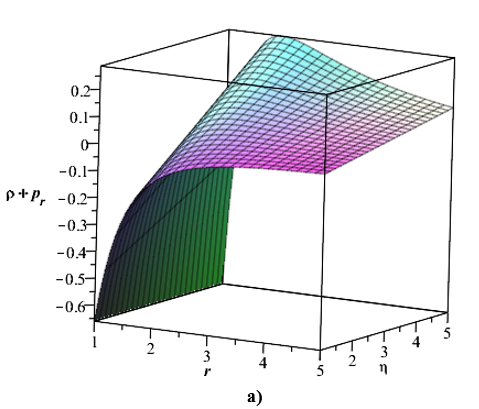}}
{\includegraphics[scale=0.48]{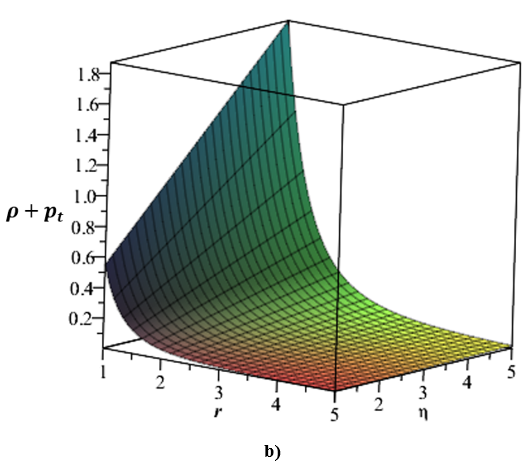}}}
 \caption{\label{fig4} a) Validity region of the NEC, $\mathrm{\rho} + \mathrm{p}_r \leq 0$, for $r_0 = 1, \mathrm{m} =3, \mathcal{B} = 4$ and $\lambda = -12$. b) \label{fig5} Validity region of the NEC, $\mathrm{\rho} + \mathrm{p}_t \geq 0$, for $r_0 = 1, \mathrm{m} =3, \mathcal{B} = 4$ and $\lambda = -12$.}
\end{center}
\end{figure}

DEC is given by
\begin{eqnarray}
  \mathrm{\rho}- \mathrm{p}_r &=& \frac{-2(\eta-1)(\mathcal{B}r^{(\mathrm{m}+1)}+1)+ \exp(c- \frac{1}{\mathrm{m}\mathcal{B}r^\mathrm{m}})(1+ \mathcal{B}r^\mathrm{m})}{\mathcal{B}r^{(\mathrm{m}+3)}(8\pi_\mathcal{F} +2\lambda)},\\\label{Eq.59}
\mathrm{\rho}- \mathrm{p}_t &=& \frac{-2(\eta-1)\mathcal{B}r^{(\mathrm{m}+1)} + \exp(c- \frac{1}{\mathrm{m}\mathcal{B}r^\mathrm{m}})(3- \mathcal{B}r^\mathrm{m})}{2(\mathcal{B}r^{(\mathrm{m}+3)})(8\pi_\mathcal{F} +2\lambda)}.\label{Eq.60}
\end{eqnarray}

\begin{figure}[hpbt]
\begin{center}
\mbox{{\includegraphics[scale=0.52]{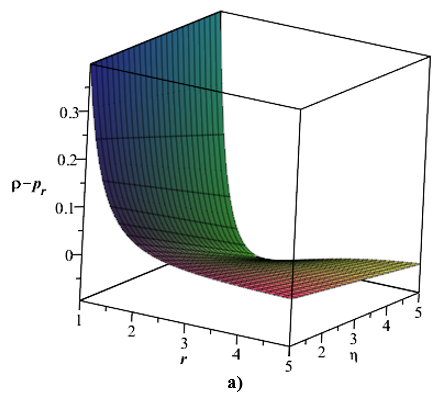}}
{\includegraphics[scale=0.49]{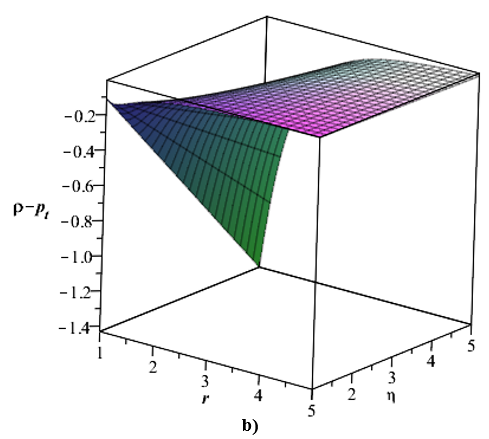}}}
 \caption{\label{fig6} a) Validity region of the DEC, $\rho \geq |\mathrm{p}_r|$, b) \label{fig7} Validity region of the DEC, $\rho \geq |\mathrm{p}_t|$, for $r_0 = 1, \mathrm{m} =3, \mathcal{B} = 4$ and $\lambda = -12$.}
\end{center}
\end{figure}

\section{Discussions and Results}

In the present theory, we have studied the SWH models with $f(\mathcal{R}, \mathcal{T})$ gravitation theory with the context of Finsler geometry using polynomial shape function and discovered a few interesting characteristics. Here, we mainly focused on $f(\mathcal{R}, \mathcal{T}) = \mathcal{R} + 2f(\mathcal{T})$, where $f(\mathcal{T}) = \lambda{\mathcal{T}}$ and $\lambda$ is parameter of constant value. The uncertainty of $f(\mathcal{R}, \mathcal{T})$ gravity WHs will depend on the range of choice of parameter. Regarding this, many researchers have been working on $f(\mathcal{R}, \mathcal{T})$ MGT, among them H. M. Manjunatha et al. \cite{Manju} (2022) investigated the models of WH in the $f(\mathcal{R})$ gravity with Finslerian approach with exponential type shape function by taking $\lambda$ = 0. In Ref.  \cite{Manju1} (2024), authors discussed WH models using an exponential shape function in the perspective of Finsler geometry with $f(\mathcal{R}, \mathcal{T})$ MGT by taking $\lambda = -12.5$. \\

In the previous section, we have plotted the embedded 2-D graph for Finslerian WH Fig. \ref{fig22}. Now, we will analyze our results by comparing them with the work done by Manjunath Malligawad \cite{Manju1}.
\begin{figure}[hpbt]
\begin{center}
\includegraphics[scale=0.4]{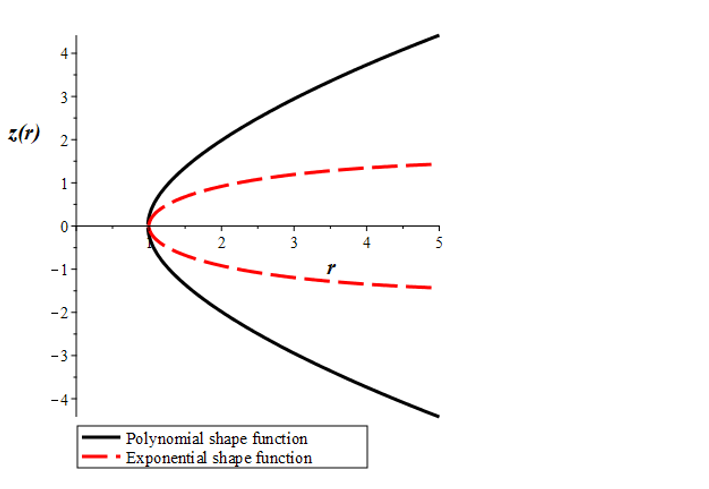}
 \caption{\label{fig24}Comparison of polynomial and exponential shape functions with $r_0 = 1$.}
\end{center}
\end{figure}

\begin{figure}[hpbt]
\begin{center}
\mbox{{\includegraphics[scale=0.33]{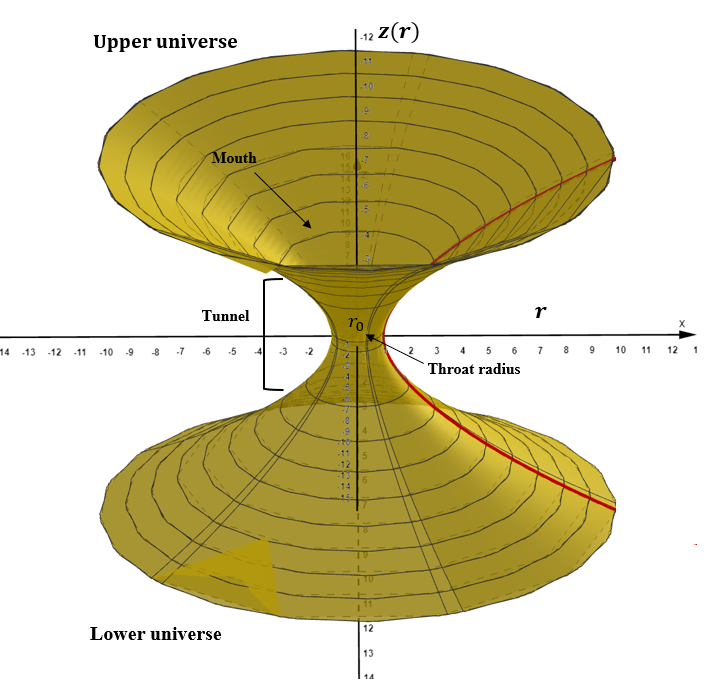}}
{\includegraphics[scale=0.23]{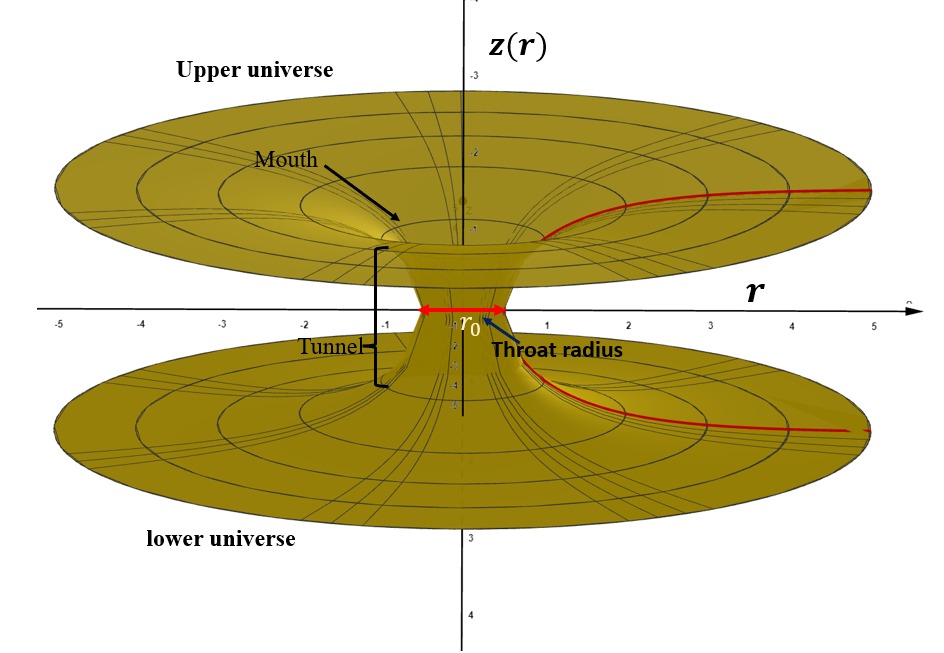}}}
 \caption{\label{fig18a} (left) Embedded 3-D WH plot for the Polynomial shape function, (right)\label{fig25} Embedded 3-D WH plot for the Exponential shape function.}
\end{center}
\end{figure}

In the  Fig. \ref{fig24} one can spot the difference in the shape of the two various kinds of the shape function $\mathrm{b(r)} = r_0e^{1-\frac{r}{r_0}}$  and $\mathrm{b(r)} = (1- \eta)r + \mathrm{A}r^{1+2\mathrm{n}}$ respectively with throat value $r_0 = 1$ for both function. We can see the bending of the exponential curve earlier than the polynomial shape function, which means the flatness of the wormhole for the exponential shape function appears to be larger with a wide range of radius shown in Fig. \ref{fig18a} (right) compared to that of the polynomial shape function Fig. \ref{fig18a} (left)\\

\begin{table}[htbp]
\begin{center}
\caption{Summary of the results of energy conditions at $r_0=1$, $\eta\geq0 ~ (0\leq\eta\leq 1)$  and $r\geq 0 ~ (0\leq r\leq 25)$ for distinct $\lambda$ values for two models.}\label{tbl1}
			\begin{tabular}{l c c }
		\toprule
		Energy    & Terms   & $\lambda =-12$  \\
		conditions & &  \\
		\midrule
		& $\mathrm{p}_r$ & $\leq 0$  \\
		
		& $\mathrm{p}_t$ & $\geq 0$  \\
		
		WEC    & $\rho$ & $\leq 0$ \\
		\midrule

        DEC      & $\rho-|\mathrm{p}_r|$ & $\leq 0 $ \\
		& $\rho-|\mathrm{p}_t|$ & $\leq 0 $ \\
        \midrule

		NEC     & $\rho+\mathrm{p}_r$ & $\leq 0$ \\
		& $\rho+\mathrm{p}_t$ & $\leq 0$ \\	
		\midrule
		
		SEC     & $\rho+\mathrm{p}_r+2 \mathrm{p}_t$ & $ =0 $ \\
		& $\rho-\mathrm{p}_r-2\mathrm{p}_t$ & $\leq 0 $ \\
		\bottomrule
	\end{tabular}

\end{center}
\end{table}

\section{Conclusions}

In the current article, we have constructed various models of SWHs that match with the $f(\mathcal{R}, \mathcal{T})$ gravitation theory within the Finslerian approach. We will discuss our conclusions for the material and geometrical content of those in this part. By the GR theory, WHs are stuffed with matter that is completely distinct from ordinary matter, known as exotic matter, and has negative mass. Many researchers have found that exotic matter is needful in studying the violation of various modified gravitation theories that account for the energy violation conditions through the effective energy-momentum tensor. As we discussed in the last sections, throat condition, i.e., $\mathrm{b}(r_0) = r_0$  at $r = r_0$, the flaring condition  $\mathrm{b}^\prime(r_0) < 1$ and asymptotic flatness that is necessary to illustrate the solutions of  WH, is obeyed in each model that we have constructed.\\

Furthermore, the redshift function has been assumed to be constant ($\mathfrak{B(r)}$ = const), which means that the hypothetical traveler's experience of tidal gravitational force is negligible. Considering the famous article by M. S. Morris and K. S. Thorne \cite{Thorne}, the authors have discovered, in a SWH, $\rho \sim r^{-2}$, and we can observe that the proportionality for $r$ when $\omega \rightarrow 1$ is predicted by our solution for the $\rho$ in the Case I of Model 2 with the similar precision. In this instance, however, our results for $\rho$ are consistent with that for the Morris-Thorne WHs along with cosmological constant Ref. \cite{Lemos} and the WHs minimally violating the NEC \cite{Lopez}.\\

And we discussed NEC and DEC in the Figs. \ref{fig15}, \ref{fig3}, \ref{fig4}, and \ref{fig7}. The study of the traversable WH's geometry has revealed a violation of NEC at WH's throat. Thus, NEC's violation may confirm the presence of exotic matter at the WH throat, which is the fundamental requirement for the existence of traversable WH. In our present model, at the throat, weaker inequality $\rho (r_0) + \mathrm{p}_r(r_0) \leq 0$ holds, which intimates the violation of NEC. Authors in \cite{Nandi} derived validity of the NEC, $\rho + \mathrm{p}_r \geq 0$, by assuming the negative energy density. Besides, SEC is also holding for both the cases of Model 2, i.e., $\mathrm{\rho} + \mathrm{p}_r + 2\mathrm{p}_t = 0 $, as one can check Eqs. (\ref{Eq.27}-\ref{Eq.29}). Violation of energy conditions shown in Table \ref{tbl1} this  violation provides strong evidence for the existence of the exotic matter in WH throat and other senses of WHs. \\

Finally, we'd like to add that we adopted a direct and accurate construction for our computations. We have achieved a detailed set of analytical solutions. And the 3-D embedded visualization of WH plot \ref{fig18a} (left) for shape function Eq. (\ref{Eq.31}) can conclude that our polynomial Finslerain $f(\mathcal{R}, \mathcal{T})$ gravity WH model is physically valid. A similar method can be incorporated in the same scenarios for different alternative gravity theories.

\section*{Appendix}\label{secA1}

We have constructed embedded 2-D and 3-D diagrams for the shape function Eq. (\ref{Eq.31}) for better visualization of the WH. We used an equatorial plane $\theta = \frac{\pi}{2}$ at a fixed time or $t$ = const, and $\eta = 1$, from these conditions Eq. (\ref{Eq.13}) reduce into the form
\begin{equation}\label{Eq.61}
  \mathcal{F}^2 = -\left(1- \frac{\mathrm{b}(r)}{r}\right)^{-1} dr^2 - r^2d\phi{^2},
\end{equation}
the above equation can be written in cylindrical co-ordinates as
\begin{equation}\label{Eq.62}
  \mathcal{F}^2 = -dz^2 - dr^2 - r^2d\phi{^2}.
\end{equation}
$z = z(r)$ represents the embedded surface in 3-dim Euclidean space. We can rewrite Eq. (\ref{Eq.62}) as
\begin{equation}\label{Eq.63}
   \mathcal{F}^2 = -\left(1+ \left(\frac{dz}{dr}\right)^2\right)dr^2 - r^2d\phi^2.
\end{equation}
Now comparing Eq. (\ref{Eq.61}) and Eq. (\ref{Eq.63}) we have
\begin{equation}\label{Eq.64}
  \frac{dz}{dr} = \pm\sqrt{\left(1- \frac{\mathrm{b}(r)}{r}\right)^{-1} - 1}.
\end{equation}
Using Eq. (\ref{Eq.64}), we plotted the embedded surface of the WH.


\begin{thebibliography}{55}
	
	\bibitem{Rosen}
 Einstein, A.; Rosen, N. The particle problem in the general theory of relativity. \textit{Phys. Rev.} \textbf{1935}, 48, 73-77.

\bibitem{Flamm}
Flamm, L. Black Holes and Wormholes - The Physics of the Universe. \textit{Phys. Z.} \textbf{1916}, 17, 448.

\bibitem{Wheel}
 Misner, C. W.; Wheeler, J. A. Classical physics as geometry. \textit{Ann. Phys. (N.Y.)} \textbf{1957}, 2, 525.

\bibitem{Bhawal}
 Bhawal, B.; Kar, S. Lorentzian wormholes in Einstein-Gauss-Bonnet theory. \textit{Phys. Rev. D} \textbf{1992}, 46, 2464.

\bibitem{Maeda}
Maeda,  H.; Nozawa, M. Static and symmetric wormholes respecting energy conditions in EinsteinGauss-Bonnet gravity. \textit{Phys. Rev. D} \textbf{2008}, 78, 024005.

\bibitem{Thorne}
 Morris, M. S.;  Thorne, K. S. Wormholes in spacetime and their use for interstellar travel: A tool for teaching general relativity. \textit{Am. J. Phys.} \textbf{1988}, 56, 395.

\bibitem{Jusu}
Jusufi, K.; Channuie, P.; Jamil,  M. Traversable wormholes supported by GUP corrected Casimir energy. \textit{Eur. Phys. J. C} \textbf{2020}, 80, 127.

\bibitem{Rich}
Richarte, M. G.; Salako, I. G.; Morais Graca, J. P.; Moradpour, H.; Ovgun, A.  Relativistic Bose-Einstein condensates thin-shell wormholes. \textit{Phys. Rev. D} \textbf{2017}, 96, 084022.

\bibitem{Hali}
 Halilsoy, M.; Ovgun, A.; Habib Mazharimousavi,  S. Thin-shell wormholes from the regular Hayward black hole. \textit{Eur. Phys. J. C} \textbf{2014}, 74, 2796.

\bibitem{BHATTI}
Bhatti, M. Z.; Yousaf, Z.; Ilyas, M. Existence of wormhole solutions and energy conditions in $f(R, T)$ gravity. \textit{J. Astrophys. Astr.} \textbf{2009}, 04, 019.

\bibitem{Harko}
 Harko, T.; Lobo, F. S. N.; Nojiri, S.; Odintsov, S. D. $f(R, T)$ gravity. \textit{Phys. Rev. D} \textbf{2011}, 84, 024020.

\bibitem{Momeni}
 Momeni, D.; Moraes, P. H. R. S.; Myrzakulov, R. Generalized second law of thermodynamics in $f(R, T)$ theory of gravity. \textit{Astrophys. Space Sci.} \textbf{2016}, 361, 228.


\bibitem{Zubair}
Noureen, I.; Zubair, M.; Bhatti, A. A.; Abbas, G. Shear-free condition and dynamical instability in $f(R, T)$ gravity. \textit{Eur. Phys. J. C} \textbf{2015}, 75, 323.

\bibitem{Azizi}
 Azizi, T. Wormhole geometries in $f(R, T)$ gravity. \textit{Int. J. Theor. Phys.} \textbf{2013}, 52, 3486.

\bibitem{Waheed}
Zubair, M.; Waheed, S.; Ahmad, Y. Static spherically symmetric wormholes in $f(R, T)$ gravity. \textit{Eur. Phys. J. C} \textbf{2016}, 76, 444.


\bibitem{Bao2}
Bao, D.; Chern, S. S.; Shen, Z. An Introduction to Riemann-Finsler Geometry, Springer \textbf{2000}.

\bibitem{Roxburgh}
 Roxburgh, I. W. Finsler spaces with Riemannian geodesics. \textit{Gen. Relativ. Grav.} \textbf{1991}, 23, 1071-1080.

\bibitem{Nekouee}
Nekouee, Z.; Narasimhamurthy, S. K.; Manjunatha, H. M.; Srivastava, S. K. Finsler–Randers model for anisotropic constant-roll inflation.  \textit{Eur. Phys. J. Plus} \textbf{2022}, 137, 1388.

\bibitem{DESY}
Pfeifer, C. The Finsler spacetime framework: backgrounds for physics beyond metric geometry, DESY-THESIS, \textbf{2013}.

\bibitem{Lorek}
Lämmerzahl, C.; Lorek, D.; Dittus, H. Confronting Finsler space–time with experiment. \textit{Gen. Relativ. Gravit.} \textbf{2009}, 41, 1345-1353.

\bibitem{Vacaru}
Bubuianu, L.; Vacaru, S. I. Axiomatic formulations of modified gravity theories with nonlinear dispersion relations and Finsler–Lagrange–Hamilton geometry. \textit{Eur. Phys. J. C} \textbf{2018}, 78, 969.

\bibitem{Vaca}
Vacaru, S. I. Principles of Einstein–Finsler gravity and perspectives in modern cosmology. \textit{Int. J. Mod. Phys. D} \textbf{2012}, 21, 1250072.

\bibitem{Xli}
Chang,  Z.; Li, X. Modified Newton's gravity in Finsler Space as a possible alternative to dark matter hypothesis. \textit{Phys. lett. B} \textbf{2008}, 668, 453-456.

\bibitem{Cartan}
Cartan, E. Les Espaces de Finsler, Actualite Scientifiques et Industrielles. Paris, Hermann \textbf{1934}.

\bibitem{Horvath}
 Horvath, J. I. A Geometrical Model for the Unified Theory of Physical Fields. \textit{Phys. Rev.} \textbf{1950}, 80, 901.

\bibitem{FRahman}
Rahaman, F.; Paul, N.; Banerjee, A.; De, S. S.; Ray, S.; Usmani, A. A.  The Finslerian wormhole models \textit{Eur. Phys. J. C} \textbf{2016}, 76, 246.

\bibitem{Ray}
Rahaman, F.; Paul, N.; De, S. S.; Ray, S.; Md. Abdul Kayum Jafry, The Finslerian compact star model \textit{Eur. Phys. J. C} \textbf{2015}, 75, 564.

\bibitem{FRahman1}
Singh, K.; Rahaman, F.; Deb, D.; Maurya, S. K. Traversable Finslerian wormholes supported by phantom energy. \textit{Front. Phys.} \textbf{2023}, 10, 1038905.

\bibitem{Manju}
Manjunatha, H. M.; Narasimhamurthy, S. K. The wormhole model with an exponential shape function in the Finslerian framework. \textit{Chin. J. Phys.} \textbf{2022}, 77, 1561-1578.

\bibitem{Manju1}
Manjunath Malligawad; Narasimhamurthy, S. K.; Nekouee, Z.; Kumbar, M. Y. Finslerian wormhole solution in the framework of modified gravity. \textit{Phys. Scr.} \textbf{2024}, 99, 045206.

\bibitem{Chang}
Li, X.; Chang, Z. Exact solution of vacuum field equation in Finsler spacetime. \textit{Phys. Rev. D} \textbf{2014}, 90, 064049.

\bibitem{Cataldo}
Cataldo, M.; Meza, P.; Minning, P. $N$-dimensional static and evolving Lorentzian wormholes with a cosmological constant. \textit{Phys. Rev. D} \textbf{2011}, 83, 044050.

\bibitem{aman}
Rahaman, F.; Kalam, M.; Sarker, M.; Ghosh, A.; Raychaudhuri, B. Wormhole with varying cosmological constant. \textit{Gen. Relativ. Gravit.} \textbf{2007}, 39, 145-151.

\bibitem{Wang}
Wang, H. C. On Finsler Spaces with Completely Integrable Equations of Killing. \textit{J. Lond. Math. Soc.} \textbf{1947}, s1-22(1), 5-9.

\bibitem{Chowdhury}
Roy Chowdhury, S.; Deb, D.; Rahaman, F.; Ray, S.; Guha, B. K. Anisotropic strange star inspired by Finsler geometry. \textit{Int. J. Mod. Phys. D} \textbf{2020}, 29, 2050001.

\bibitem{Chang2}
Chang, Z.; Li, X. Lorentz invariance violation and symmetry in Randers–Finsler spaces. \textit{Phys. Lett. B} \textbf{2008}, 663, 103-106.

\bibitem{Chang1}
Li, X.; Wang. S.; Chang. Z. Finslerian Perturbation for the $\lambda$CDM Model, \textit{Commun. Theor. Phys.} \textbf{2014}, 61, 781.

\bibitem{Pfeifer}
 Pfeifer, C.; Wohlfarth, M. N. R. Finsler geometric extension of Einstein gravity. \textit{Phys. Rev. D} \textbf{2012}, 85, 064009.

\bibitem{PStavrinos}
Stavrinos, P.; Vacaru, O.; Vacaru, S. I. Modified Einstein and Finsler like theories on tangent Lorentz bundles \textit{Int. J. Mod. Phys. D} \textbf{2014}, 23, 1450094.

\bibitem{saho}
 Moraes, P. H. R. S.; Sahoo, P. K. Modeling wormholes in $f(R, T)$ gravity \textit{Phys. Rev. D} \textbf{2017}, 96, 044038.

\bibitem{Lobo}
Garcia, N. M.;  Lobo, F. S. N. Wormhole geometries supported by a nonminimal curvature-matter coupling. \textit{Phys. Rev. D} \textbf{2010}, 82, 104018.

\bibitem{Hochberg}
Hochberg, D.; Visser, M. Null Energy Condition in Dynamic Wormholes. \textit{Phys. Rev. Lett.} \textbf{1998}, 81, 786.

\bibitem{Raha}
Rahaman, F.; Kalam, M.; Rahman, K. A. Conical thin shell wormhole from global monopole: A theoretical construction. \textit{Acta Phys. Polon. B}  \textbf{2009}, 40, 1575-1590.

\bibitem{Lemos}
 Lemos, J. P. S.; Lobo, F. S. N.;  Quinet de Oliveira, S. : Morris-Thorne wormholes with a cosmological constant. \textit{Phys. Rev. D} \textbf{2003}, 68, 064004.

\bibitem{Lopez}
  Bouhmadi-Lopez, M.; Lobo, F. S. N.; Martin-Moruno, P. Wormholes minimally violating the null energy condition. \textit{JCAP} \textbf{2014}, 11, 007.

\bibitem{Nandi}
Nandi, K. K.; Bhattacharjee, B.; Alam, S. M. K.; Evans, J. Brans- Dicke wormholes in tha Jordan and Einstein frames. \textit{Phys. Rev. D} \textbf{1998}, 57, 823.
	
	
\end{thebibliography}
\end{document}